\PassOptionsToPackage{unicode}{hyperref}
\PassOptionsToPackage{hyphens}{url}
\documentclass[10pt,twocolumn]{article}

\usepackage[numbers, sort&compress]{natbib}
\usepackage[margin=0.75in]{geometry}
\usepackage{times}
\usepackage{tabularx}
\usepackage{xcolor}
\usepackage{amsmath,amssymb}
\usepackage{iftex}
\usepackage{lmodern}

\usepackage{upquote}
\usepackage[]{microtype}
\usepackage{comment}
\UseMicrotypeSet[protrusion]{basicmath} 

\makeatletter
\usepackage{parskip}
\makeatother
\usepackage{longtable,booktabs,array}
\usepackage{calc} 
\usepackage{etoolbox}
\usepackage{graphicx}
\makeatletter
\newsavebox\pandoc@box

\def\fps@figure{htbp}
\makeatother
\usepackage{soul}
\usepackage{bookmark}
\usepackage{xurl}
\hypersetup{
  hidelinks,
  pdfcreator={LaTeX via pandoc}}

\title{Cognitive Dark Matter: Measuring What AI Misses}
\author{
Patrick J.\ Mineault\thanks{Amaranth Foundation} \and 
Thomas L.\ Griffiths\thanks{Princeton University} \and 
Sean Escola\thanks{Protocol Labs}}
\date{\today}  

\begin{document}

\twocolumn[
\maketitle
\begin{quotation}\noindent\small
We propose that the jagged intelligence landscape of modern AI systems arises from a missing training signal that we call ``cognitive dark matter'' (CDM): brain functions that meaningfully shape behavior yet are hard to infer from behavior alone. We identify key CDM domains---metacognition, cognitive flexibility, lifelong learning, abductive reasoning, social and common-sense reasoning, and emotional intelligence---and present evidence CDM-loaded functions are largely unmeasured in current AI benchmarks, and that large-scale neuroscience training datasets which could be used to instill these capabilities do not yet exist. We then outline a research program centered on three complementary data types designed to surface CDM for model training: (i) latent variables from large-scale cognitive models, (ii) process-tracing data such as eye-tracking and think-aloud protocols, and (iii) paired neural--behavioral data. These data will enable AI training on cognitive process rather than behavioral outcome alone, producing models with more general, less jagged intelligence. As a dual benefit, the same data will advance our understanding of human intelligence itself.
\end{quotation}
\vspace{1.5em}
]
\renewcommand{\thefootnote}{\fnsymbol{footnote}}
\footnotetext[1]{Amaranth Foundation}
\footnotetext[2]{Princeton University}
\footnotetext[3]{Protocol Labs}
\renewcommand{\thefootnote}{\arabic{footnote}}
\setcounter{footnote}{0}

Over the past several years, AI has made remarkable strides
\cite{Krizhevsky2012-uz,Silver2018-qz,Abramson2024-wd,Brown2020-bk}.
Benchmarks, where they exist, have increasingly become saturated: while
it took close to 20 years to go from defining handwriting and speech
recognition benchmarks to solving them at a human level, more recent
benchmarks focusing on natural language processing and advanced
reasoning have reached saturation within a period of around 2 years
\cite{Kiela2021-oi}. The potent combination of clear
definitions of success, as captured by benchmarks, and unleashing
impressive amounts of compute on large-scale datasets has led to rapid
advances, first in vision and language, and now in other domains,
including, in particular, software programming, mathematics, and
biology.

Nevertheless, there remain clear gaps in the cognitive capabilities of AI systems.
Artificial intelligence is \emph{jagged}
\cite{Dell-Acqua2023-uu, pacchiardi2025framework, morris2026characterizing}: while AI systems are
remarkably adept at some tasks that are very difficult for individual
humans, they often stumble on seemingly innocuous tasks. They suffer from a surprising lack of generalization \cite{mancoridis2025potemkinunderstandinglargelanguage, rane2026investigatingconceptalignmentusing}, which can be established by designing control conditions that probe cognitive abilities adversarially \cite{rane2025position}. It can be
difficult for a human \cite{vafa2024largelanguagemodelsperform} to predict the boundary between what is feasible and
infeasible for an AI model, although progress has been notable in building machine learning systems that anticipate the success of other machine learning systems \cite{zhou2026predictable, zhou2026general}. 

Why is performance jagged? Our thesis is
that progress concentrates in capabilities where training data and evaluation metrics are
rich, and stalls where they are thin. Because the behavior of an AI system reflects its training data, cognitive abilities that are under-represented in that data will be unreliable in the resulting system; the jagged profile we observe is the downstream signature of which capabilities the data does and does not carry. We call human brain functions that are not captured in the current AI paradigm 
\textbf{cognitive dark matter} (CDM): capabilities that materially shape
intelligent behavior yet are under-represented in standard AI training data, and thus are poorly instilled in models during training. While we are not the first to use the term dark matter to refer to hidden cognitive processes \cite{ackerman2000domain, zhu2020dark, bolotta2022social}, we adopt it as an evocative metaphor for this phenomenon. Cosmological dark matter accounts for much of the matter in the universe, and it is not visible. Much as the existence
of cosmological dark matter is inferred from its macro-effects---heavy
dark matter must be abundant for galaxies to hold together under gravity---the existence of CDM is inferred from the fact that humans \emph{must}
have certain competencies in order to survive and thrive, but these have
proved difficult to elicit from and evaluate in behavior. Furthermore, this CDM is under-represented in the benchmarks that frontier labs have adopted to evaluate and optimize their models, so that the resulting deficits often escape detection. 

The data used to train current AI models largely amounts to human
\emph{behavior}: the text, images, and other digital traces that we
generate. Underlying these data is a set of complex causal processes:
human \emph{cognition}. The thoughts and feelings that result in our
behavior are latent variables that AI models rarely have access to. What
we are calling cognitive dark matter is largely made up of these latent
variables. By giving AI models better access to the underlying causal
processes, we can potentially train those models to perform in ways that
are more similar to humans, more interpretable, and more generalizable.

Here, we sketch a working definition of CDM, argue that its
under-measurement explains jagged intelligence, and propose a concrete
path forward: using rich data derived from human minds and brains to
supervise models on process, not just outcomes.

\section{Jagged intelligence and its sources}\label{an-example-of-jagged-intelligence-and-missing-cdm-in-ai}

\begin{figure}[htbp]
  \centering
  \includegraphics[width=.9\columnwidth]{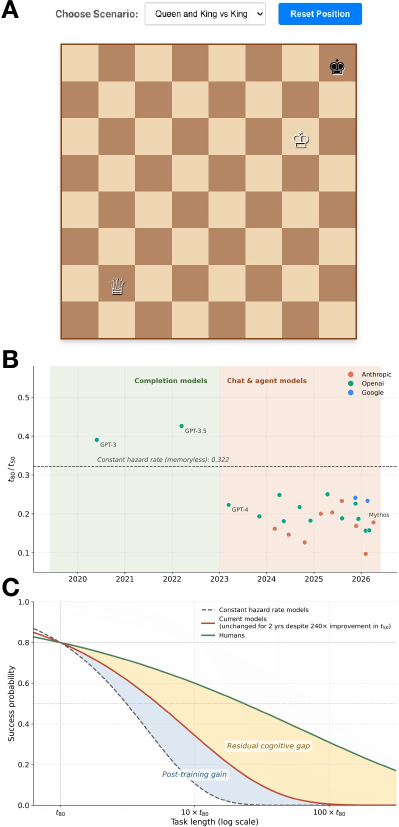}
  \caption{
  Multistep task performance.
  \textbf{A}. Mate-in-one app built by Claude Code. White to play and notice that the position is invalid;
  black is already in check. This could be avoided by checking the
  validity of the proposed positions via the already imported libraries. Failing to identify one's cognitive weaknesses is a failure of
  metacognition.
  \textbf{B}. $t_{80}/t_{50}$ ratios for leading frontier models \cite{METR}. Following the introduction of extensive post-training pipelines with GPT-4, models demonstrated better-than-constant hazard rates, though still worse than human. Human $t_{80}/t_{50}$ = 0.04 (not shown; \cite{Kwa2025-rf}).
  \textbf{C}. Survival curves with $t_{80}/t_{50}$ ratios of 0.32 (for constant hazard rate models, dotted black line), 0.2 (GPT-4 and later models, red line), and 0.04 (humans, green line). The shaded blue region shows the cognitive benefit from post-training. The shaded yellow region is the residual cognitive gap with humans that hasn't closed since GPT-4.
  }
  \label{fig:chess-endgame}
\end{figure}

Despite their ability to produce sophisticated solutions to many problems, current AI systems fail surprisingly on others. Consider a well-defined software engineering task: asking an AI system to
create a web application that allows one to practice 3 chess endgames
(mate-in-one) of its choice. In our tests with state-of-the-art models
at the time of writing\footnote{This particular issue might be fixed by
  the time you read these lines, but the general point stands: it's hard
  to predict what models fail at and why.} (GPT-5.1-Thinking, Claude
Opus 4.5, Gemini 3.0 Pro), we have not been able to reliably complete
this task in a single prompt (Figure \ref{fig:chess-endgame}a). The reason is rather
unintuitive: most frequently, models generated \emph{invalid endgames}.

On the one hand, it is remarkable that these models have encyclopedic
knowledge of software packages and can write competent code to display a
chess board with draggable pieces, involving hundreds of valid lines of
HTML, CSS, and JavaScript. A novice programmer might struggle with a
drag-and-drop implementation, but would at least verify that the app
they've created actually accomplishes the overall goal: practicing
endgames.

This tendency to surprisingly omit key components of a complex task is not limited to writing chess apps. We have seen remarkable improvements in the ability of coding agents to
complete long tasks, with a doubling of task length every 7 months
\cite{Kwa2025-rf}. State-of-the-art models now complete,
with 50\% probability, tasks that would take a human over half a day \cite{METR2025-oz}. However, the mechanism by
which they've reached this capability appears quite different from
humans. Ord \cite{Ord2025-wy} noted that the successes
and failures of AI systems on these coding tasks are consistent with a constant hazard
rate over time. That is, if a task contains $n$ subtasks and the
probability of success for each subtask is $p$, the probability of
success for the overall task is $p^n$. This is problematic: unless $p = 1$, an
exponential falloff will doom long tasks
\cite{Dziri2023-jo,LeCun2025-ky}. 

Given recent rapid progress -- e.g., with Anthropic's Mythos showing a 50\% completion rate for tasks up to 16 hours \cite{METR} -- we revisited Ord's analysis and estimated the survival curves of frontier models from their $t_{80}/t_{50}$ ratios (i.e., the ratios of human performance durations for tasks that models complete with 80\% and 50\% success rates). For survival curves with constant hazard rates, this ratio should be about 0.3, which is indeed what is seen with GPT-3 and GPT-3.5. However, starting with GPT-4 and continuing through Mythos, $t_{80}/t_{50}$ ratios are instead about 0.2, implying a non-constant decreasing hazard rate (Figure \ref{fig:chess-endgame}b). This is a meaningful improvement, albeit far from human performance (with a $t_{80}/t_{50}$ ratio of 0.04 \cite{Kwa2025-rf}). We hypothesize this is a consequence of the substantially expanded post-training pipeline which was introduced with GPT-4 \cite{OpenAI2023-qq} and has continued into the present -- i.e., post-training is able to recover some but not all of the CDM functions (described below) that humans rely on when executing multistep tasks (Figure \ref{fig:chess-endgame}c).

Remarkably, $t_{50}$ has increased 240 times from GPT-4 (4 minutes) to Mythos (16 hours) \cite{METR} despite no change in the survival curve shapes (as measured by $t_{80}/t_{50}$). Essentially, recent frontier model performance improvements have resulted from stretching survival curves by decreasing per-step error rates proportionally (e.g., through the use of verified tools, cf. \cite{schick2023toolformer}), rather than changing the fundamental dynamics of multistep
task completion. Unless the persistently short tails of the survival curves of AI models versus humans are eliminated, trustworthy model performance on long-horizon tasks may remain elusive.

There are multiple well-documented mechanisms by which humans can
display significantly better-than-constant hazard rates, at least until boredom or
fatigue kick in \cite{Mackworth1948-mh}. These include
learning within a task \cite{Kotovsky1985-pz}, insight
\cite{Metcalfe1987-tu}, or incubation
\cite{Sio2009-dz}. More broadly,
\emph{\textbf{metacognition}} \cite{Fleming2024-ll} -- the self-reflective monitoring of ongoing performance and, when needed,
triggering of corrective control -- endows humans with the ability to
\emph{recover from compounding errors}, even if human competence at
individual subtasks is weaker than that of an equivalent machine.
Competent adults with strong capabilities in some domains and weaknesses
in others can compensate for them: a procrastinating student re-doubles
their effort upon receiving a D on a midterm. In our chess endgame
example, metacognition appears to be missing. Claude Code has access to
feedback and tools in a loop. Thus, after multiple failures, the model
\emph{could} choose a different tack, leveraging its superhuman coding
abilities and using already imported chess libraries to check that the
proposed positions are valid. Instead, Claude Code perseveres with a bad
strategy, and so displays low \emph{\textbf{cognitive flexibility}}. Our intent here is not to claim that this particular example may not be patched in future model releases (perhaps even by publication of this article), but rather that we can connect the surprising deficits of AI models to specific cognitive abilities.


In addition to metacognition and cognitive flexibility, current AI
models display difficulty in applying common sense to reason about the world, updating their
knowledge, generating genuinely novel explanations, and understanding
the minds and feelings of humans. We presume that these missing
capabilities, as well as others, underlie jagged intelligence and impede
the transition from chatbots to full agents. More significantly, the
lack of these abilities -- the causal processes that underlie human
performance -- means that our intuitions about how the behavior of AI
systems will generalize from task to task are dramatically off the mark.

Put another way, the problem of jagged intelligence is not just one of
task failure, but of failure type. Hardcoding invalid endgames in an
otherwise sophisticated web app for chess is not merely wrong; it's
alien. AI systems that fail in unpredictable ways resist integration into human
social networks. The goal is not merely to reduce failures, but to
ensure that failures are detectable and interpretable. One path to
achieve this is recapitulating the cognitive processes that generated
human outputs. Systems built on these foundations won't just generalize
better, they will also fail better, in ways we can understand,
anticipate, and correct.

\section{Jagged intelligence stems from a lack of
training data}\label{jagged-intelligence-stems-from-a-lack-of-training-data}

Some capabilities are inherently hard to measure: they are seldom
expressed, and in idiosyncratic ways, which preclude large-scale,
repeatable data collection. For example, some are expressed only
privately, as in metacognition; are expressed only once, like insight -- a Eureka moment -- and don't leave many behavioral traces online; or are
expressed when people interact together in private social settings,
which again are only partially reflected in internet data. Behavior, 
expressed in text and other measurable outputs, is easy to measure.
The cognitive processes that produce that behavior are not.

Our thesis is that we need more training data in which these hard-to-measure
elements are made visible. A direct approach would be to measure and
replicate, in humans, the \emph{process} by which a complex task is
completed. \emph{Process supervision}
-- rewarding a model for producing not
just the desired outcome, but the intermediates leading to the correct
outcome -- can improve the performance of models on mathematical
reasoning \cite{Luo2024-rc}. However, for most tasks, this direct approach is not possible
because the ``correct'' intermediate steps are unknown. An alternative
is to surface the latent variables in human brains and minds that reflect cognitive dark matter
and that underlie the generation of behavior, and then use these latents
as supervision signals to train general-purpose models.

We see three kinds of data that can surface the content of cognitive
dark matter. Prior to introducing these, it is useful to think of a schematized generative process for human behavior in which variable $X$ (brain activity) has influence on downstream variables $Y$ (typical AI training data like text) and $Z$ (other behavioral outputs that aren't typically used for training). The training objective is to produce a model that, when sampled from, generates outputs with the same distribution as $Y$.

Our first proposal is to add to the training mix \textbf{inferred latent variables from cognitive models} fit to large-scale rich behavioral data. The idea here is that well-designed and validated cognitive models are imbued with useful and important inductive biases about human cognition. These models essentially impose a strong prior on $X$, which thus favors certain -- ideally more human-like -- solutions over others during training. The advantage of this approach is that it doesn't require new types of data other than those typically used for training, but it does require that the hypothesis space of the cognitive models that are employed cover human-like solutions.

This approach is illustrated by problem-solving:
classic work by Newell and Simon \cite{Newell1972-il}
suggested that we can think about human problem solving in terms of
searching a tree of possibilities, and identified heuristic strategies
people use to make that search efficient, an idea that has been applied succesfully in AI \cite{yao2023treethoughtsdeliberateproblem}. Modern AI systems are trained
on data that are generated by such processes, in the form of reasoning
traces that are used to fine-tune models. However, search is just one
kind of cognitive process; the pipeline from behavior to cognitive model to synthetic data could be applied to other cognitive domains. By measuring behavior when people make
complex decisions, engage in planning, and reason about the minds of
others, we can develop accurate cognitive models of those behavioral paradigms
\cite{Griffiths2024-du}, which can then be used to
develop synthetic CDM-enriched data in these domains. This requires collecting large-scale behavioral studies of
human cognition \cite{Peterson2021-ok,Almaatouq2022-ye} in specific domains.

\begin{table*}[htbp]
  \centering
  \label{tab:ai-tiers}
  \begin{tabularx}{\linewidth}{l >{\raggedright\arraybackslash}p{2.5cm} >{\raggedright\arraybackslash}X}
    \toprule
    AI Tier & Capability Level & Examples \\
    \midrule
    L1 & High (superhuman in some cases) &
      Visual Perception, Language Comprehension, Language Production,
      Face Recognition, Auditory Processing, Reflexive Responses \\
    L2 & Partial progress &
      Planning, Logical Reasoning, Decision-making, Working Memory,
      Reward Mechanisms, Multisensory Integration, Spatial Representation
      \& Mapping, Attention, Episodic Memory, Sensorimotor Coordination,
      Scene Understanding \& Visual Reasoning, Visual Attention \& Eye
      Movements, Semantic Understanding \& Context Recognition, Adaptive
      Error Correction, Motor Skill Learning, Motor Coordination \\
    L3 & Rarely explored &
      Cognitive Flexibility, Inhibitory Control, Social Reasoning \&
      Theory of Mind, Empathy, Emotional Intelligence, Self-reflection,
      Tactile Perception, Lifelong Learning, Cognitive Timing \&
      Predictive Modeling, Autonomic Regulation, Arousal \& Attention
      States, Motivational Drives \\
    \bottomrule
  \end{tabularx}
    \caption{Tiers of cognitive capabilities in AI systems, adapted from Liu et al.~\cite{Liu2025-es}. Their categorization
    stratifies different cognitive capabilities in terms of how well they are explored in AI systems.}
\end{table*}

Our second proposal is to add new kinds of behavioral data, specifically 
\textbf{process-tracing data}. Methods such as eye-tracking or
mouse-tracking \cite{Glaholt2011-ou,Maldonado2019-ow,mondal2025gaze,kerr2025eyerobotlearninglook}
can reveal what people are attending to and what options they are
considering, elucidating their cognitive processes. Triplet odd-one out
tasks reveal the implicit geometry of the manifold of images, and can
help improve performance and robustness
\cite{Muttenthaler2025-zp}. Process tracing techniques
also include ``think aloud'' paradigms, where people are encouraged to
describe the steps they are executing when working on a problem
\cite{Schulte-Mecklenbeck2019-ai}. This results in rich
text descriptions of mental processes, making ineffable cognitive
processes visible -- something that has already been shown to provide an
advantage for fine-tuning AI systems \cite{Luo2024-rc}.
By collecting large amounts of process-tracing data, we obtain
constraints and datasets that can focus AI systems on more human-like
perceptual experiences and thought processes. Essentially, this approach should sharpen the posterior over $X$ by estimating it from both $Y$ and $Z$ versus $Y$ alone, assuming that $Z$ contains new non-redundant information.

Finally, \textbf{paired neural and behavioral data} offer a unique
opportunity for training AI systems, alleviating the need to infer $X$ from behavior by measuring it directly. Indeed, multiple studies have shown
glimpses that generic models can be fine-tuned with brain data, leading
to improved performance advantages. Recent work has demonstrated that
fine-tuning audio and video models on neural activity improved
performance on downstream semantic and social tasks
\cite{Moussa2024-za,Vattikonda2025-hy,Policzer2025-eh}.
Furthermore, training on neural data can increase robustness to
adversarial stimuli
\cite{Li2019-dg,Dapello2020-af,Safarani2021-ui,Dapello2022-gg}. These ``brain-tuning'' approaches are implemented similarly to methods used for distilling from large to small models (e.g., feature-based knowledge distillation \cite{romero2015fitnetshintsdeepnets, gou2021knowledgedistillationsurvey, tian2022contrastiverepresentationdistillation}), albeit with the source ``model'' being the brain and the target being a frontier AI.
We are not the first to suggest \cite{Fong2018-sv} that
neural data could be used to directly train artificial intelligence. Our
point is more subtle, which is that neural data could be most profitably
used to train artificial intelligence systems in domains which are
heavily loaded on cognitive dark matter, which is otherwise difficult to
capture from standard training data.

\begin{figure*}[htbp]
  \centering
  \includegraphics[width=\textwidth]{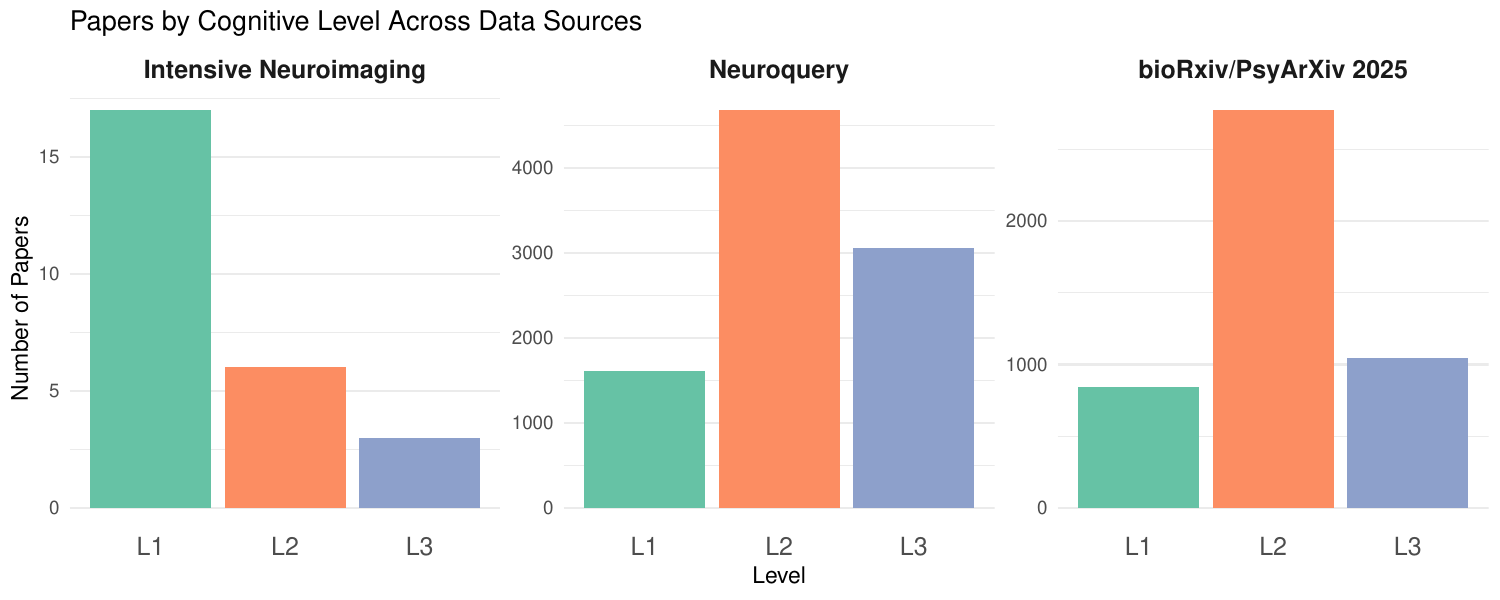}
  \caption{Distribution of cognitive capabilities 
  measured in intensive neuroimaging datasets by AI tier, and in the neuroscience
  literature at large. While the field of neuroscience as a whole is
  interested in L2 and, to a smaller extent, L3 capabilities, foundational
  large-scale datasets which could be used for training do not yet exist for these capabilities.}
  \label{fig:neuroimaging-tiers}
\end{figure*}

\section{\texorpdfstring{Charting the training data gap
}{Charting the training data gap }}\label{charting-the-training-data-gap}

Bringing this vision to life will require scaling (i)
behavioral data for robust cognitive models, (ii) process-tracing data, and (iii) paired neural-behavioral data, all in the
context of tasks that load heavily on CDM. For neural-behavioral data in
particular, the issue isn't just scale -- it is subject matter. While many taxonomies that index capabilities motivated by cognitive science have been proposed (e.g. \cite{Poldrack-2011, kahneman2011thinking, voudouris2025morgan, Hendrycks2025-el, putnoki2026cognitive}), to understand whether neural-behavioral data is relevant to AI training, we need a taxonomy that cross-tabulates those capabilities against those of AI. 
A useful taxonomy of AI agent capabilities in this regard was introduced by Liu et al. \cite{Liu2025-es}: they
define capabilities supported by human cognition and their mastery in
current AI in a three-tier classification, ranging from L1 (high
capability) to L2 (partial progress) to L3 (rarely explored) (see Table
1). L1 covers cognitive functions like language comprehension and visual perception, which have
been mastered by current AI systems, while L3 covers
functions aligned with cognitive dark matter (with L2 comprising intermediate functions).

A survey of large-scale human neural datasets reveals that intensive
neuroimaging \cite{Kupers2024-sf} has been heavily skewed towards two domains, vision and
language, which are by-and-large mastered by AI models: they are
classified as L1. This is in contrast with the broader field of
neuroimaging and neuroscience. To take stock of the current state of this field, we re-analyzed the NeuroQuery data \cite{Dockes2020-ci}, which has previously been used for meta-analyses on neuroimaging data; we complemented this dataset with the full set of neuroscience biorxiv and PsyArXiv preprints from 2025, extending the analysis beyond human imaging studies. This revealed that many studies focus on
L2 and, critically, L3 domains, but these are typically small-scale,
single-focus, hypothesis-driven, short-duration experiments. In short, we have simply not built the kind of CDM-loaded,
intensive, foundational neural-behavioral dataset that would allow translation to AI.

To summarize, with few exceptions, those aspects of cognition that are
well-mastered by AI are those that are well-measured by large-scale
neuroscience datasets. Conversely, those aspects of AI that are not yet
mastered are also poorly measured in large-scale neuroscience
experiments. This gap suggests a clear opportunity with dual benefits:
obtain large-scale neuroscience datasets during complex cognition, with
a high loading on CDM; and then use these datasets to both better understand
ineffable aspects of human cognition and fill the gaps in the jagged intelligence landscape of AI systems.

\begin{figure*}[htbp]
  \centering
  \includegraphics[width=\textwidth]{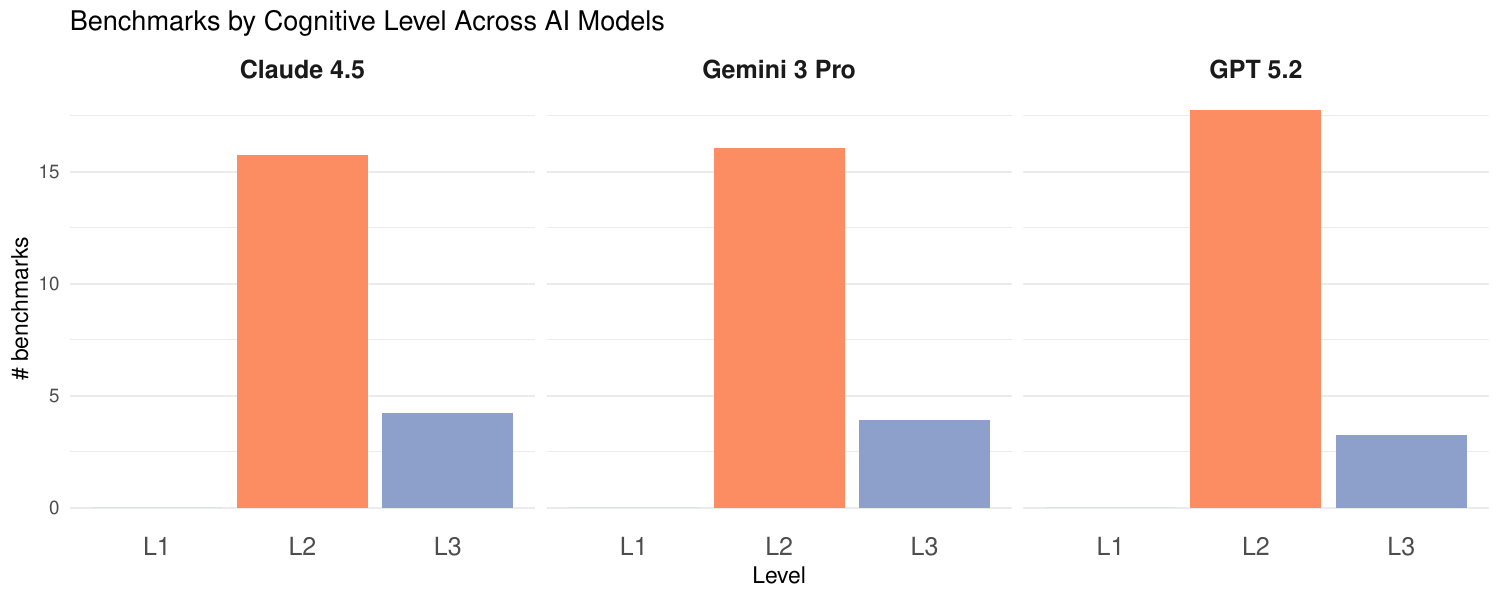}
  \caption{Distribution across AI tiers of the benchmarks used by leading
  AI labs for evaluation. We assigned AI tier labels to each benchmark reported in the
  three models' launch documents
  \cite{GDM2025-nw,Anthropic2025-ks,OpenAI2025-br}
  according to the highest AI tier of the cognitive functions the
  benchmark is designed to probe. Frontier models are almost exclusively
  benchmarked against tests that target L2 capabilities. This is an
  interesting development from 2023-era models, which were evaluated on
  L1 capabilities \cite{OpenAI2023-qq}. What used to be the frontier
  has faded into the background.}
  \label{fig:benchmark-tiers}
\end{figure*}

\section{Charting the evaluation gap}

We now turn our attention to evaluation. We focus here on the evaluation suites published by frontier labs, since these are the measurements against which commercial models are optimized and therefore the ones that most directly steer where capability gains accrue. Surveying these suites, we find a substantial underrepresentation of CDM. Categorizing each benchmark in the evaluation suites published as part
of the releases of Claude Opus 4.5
\cite{Anthropic2025-ks}, Gemini 3 Pro
\cite{GDM2025-nw}, and GPT 5.2
\cite{OpenAI2025-br} by the highest tier of AI function
(according to the Liu et al. schema) that the benchmark probes, we find
that the evaluation suites largely focus on L2 cognitive functions with
little L3 function characterization (Figure \ref{fig:benchmark-tiers}; see Methods). Two clear
exceptions are ARC-AGI-2 \cite{Chollet2025-nr} and
$\tau^2$-bench \cite{Barres2025-bq}. The former challenges
models to discover and operate under the constraints of
never-before-seen rules and thus requires \emph{cognitive flexibility},
an L3 function. The latter assesses how well models cooperate with users
in customer support environments and requires the L3 function of
\emph{social reasoning and theory of mind}.

As Figure \ref{fig:benchmark-tiers} shows, 2025-era evaluation suites don't test L1 functions in
isolation, focusing almost exclusively on L2, suggesting that the
field assumes human or super-human performance for L1 functions by all
modern models. In contrast, evaluation suites for GPT-4 included
multiple benchmarks with L1 functions as the highest AI tier
\cite{OpenAI2023-qq}. So while frontier-lab evaluation has
evolved to match the cognitive frontier of current models, assessment of
CDM-loaded L3 functions is still lacking. This results in a skewed view of performance with
the jagged edges of modern models largely hidden, compounding the well-documented difficulty in measuring AI performance accurately \cite{zhang2025remembering, eriksson2025can, raji2021ai}.

\section{What to collect next: a CDM wish
list}\label{what-to-collect-next-a-cdm-wish-list}

We believe that the time is ripe to collect training datasets that capture CDM
and to design evaluation benchmarks to properly measure it in AI. It has never been
easier to collect internet-scale behavioral data across a wide range of
complex cognitive tasks \cite{Allen2023-ub}. Intensive
neuroimaging datasets \cite{Kupers2024-sf} have provided
a blueprint for collecting these datasets in the context of
neuroimaging; in addition to fMRI, OPM-MEG and functional ultrasound
imaging can help us peer into the brain non-invasively; and invasive
recording technologies including human Neuropixels
\cite{Paulk2022-yj} and recordings during epilepsy
monitoring \cite{Peterson2022-id} give increasing access
to fine-grained neural activity that has otherwise eluded us.

We propose that longitudinal, large-scale, high-entropy datasets should
be collected in humans with a high loading on CDM capabilities that we
suspect could be used as training data to help solve jagged intelligence:

\begin{itemize}
\item
  \emph{\textbf{Metacognition}}: Humans have limited cognitive resources and
  must decide how to deploy those resources effectively
  \cite{Lieder2019-ot}. This requires an interconnected
  set of metacognitive abilities: self-assessment, to know what we can
  and can't do \cite{Fleming2017-ta}; task monitoring,
  to track progress against our expectations
  \cite{Ackerman2017-qv}; and meta-reasoning, deciding
  which strategies to pursue in new tasks
  \cite{Lieder2017-sq}. These abilities must be
  calibrated: if we are systematically over-confident, we take on tasks
  where we lack good prospects for success, or pursue overly simplistic
  strategies for complex tasks. While humans consistently make
  calibration errors, those errors are often explicable as rational
  inferences from limited data \cite{Moore2008-vs}. By
  contrast, current AI models consistently overestimate their abilities
  \cite{Xiong2023-vx,OpenAI2023-qq,Kalai2025-km,Johnson_2026,Wang_2025,ackerman2025evidence,li2026towards}.
\item
  \emph{\textbf{Cognitive flexibility}}: A hallmark of biological intelligence
  is the ability to quickly develop adaptive strategies in novel
  environments or in response to feedback. This capability is
  demonstrated most classically in the Stroop task, where a switch in
  rules requires a new behavioral strategy and the suppression of a
  prior one \cite{Golden2012-eh}. This is not a strictly
  human capability with rodent and monkey versions of the Stroop task
  demonstrating that other mammals can infer context switches and apply
  context-dependent strategies
  \cite{Haddon2006-du,Washburn1994-kp}. Models, on the
  other hand, show inappropriate perseverative behavior  \cite{kennedy2024cognitive,Goto_2025,moment2025triangulating}, as in our chess
  example, where failing strategies were repeatedly applied despite
  clear feedback. The ARC-AGI series of benchmarks also demonstrates
  this biological versus artificial intelligence disparity. When faced with
  novel situations with unknown rules, humans quickly figure out what to
  do while models struggle even with considerable computational
  resources \cite{Chollet2025-nr}.
\item
  \emph{\textbf{Lifelong learning}}: Humans learn to see, hear, and control
  their bodies during development despite limited experience
  \cite{Lake2015-kb}; teenagers learn to drive in a
  dozen hours; and we continue to adapt our behavior over a lifetime,
  despite changes in our environment, our bodies, and our senses.
  Current models require retraining on massive datasets to incorporate
  new information \cite{lin2021the,wang2023trace,Ishibuchi_2023}. Continual and data-efficient learning could unlock AI
  that acquires and integrates knowledge and skills rapidly, overcomes
  catastrophic forgetting, and avoids pitfalls like the reversal curse
  \cite{Berglund2023-zl}.
\item
  \emph{\textbf{Abductive reasoning}}: AI research has focused heavily on improving
  deductive reasoning, using reinforcement learning to support the
  generation of chains of thought that result in verifiable outcomes in
  settings like mathematics and logic \cite{Wei2022-os}.
  Large language models also show some success in inductive reasoning:
  training for next-token prediction implicitly trains these systems to
  perform Bayesian inference with appropriate priors, and their behavior
  can be analyzed as such
  \cite{Griffiths2023-nz,Xie2021-yn}. However, there is
  a third kind of reasoning that is much harder to capture in datasets
  because of its intrinsic rarity: abductive reasoning, in which we
  leap from observations to a potentially novel explanation
  \cite{Walton2014-qm, Del_2023,Thagard_2025,he2025gear}. Abduction is the kind of insight
  that leads a scientist to a new discovery, or an engineer to a new
  design---recognizing a kind of structure in the data that hadn't
  previously been posited. Capturing this human capacity is a key
  challenge for creating AI systems that can act as effective scientists
  capable of making novel discoveries or transcending the information in
  their training data.
\item
  \emph{\textbf{Social and common-sense reasoning}}: Human behavior relies on
  mental models of the world: causal relationships, governing rules,
  dynamical evolution over time. These range from intuitive physics---what happens when the cup is pushed off the table
  \cite{McCloskey1983-ir}?---to complex multidegree
  social interactions---what will she think that he thinks that she
  thought when he said, etc.~\cite{Kinderman1998-co}?
  Accurate (enough) mental models permit reasoning and planning in novel
  scenarios: the essence of generalization. Current AI develops world
  models that are at best incomplete and at worst wrong
  \cite{Vafa2024-jd}, with demonstrated resultant
  cognitive impairment \cite{wu2025how,xu2025socialmaze,Wang_2024}.
\item
  \emph{\textbf{Emotional intelligence}}: Emotional intelligence involves the
  recognition of affective states in oneself and others, the influence
  of those states over behavior, and their modulation for contextually
  adaptive behavior \cite{Brackett2006-aj}. Examples of
  emotionally-guided behavior include risk perception shifts in the
  presence of fear or anger
  \cite{Lerner2015-tb,Loewenstein2003-pd}. We can
  sidestep the anthropomorphic framing of whether or not machines can
  ``feel'' and still assess affective cue perception, emotional
  appraisals, and the prosociality of responses in emotionally-laden
  contexts
  \cite{Barrett2016-xl,Picard2000-qq,Singer2009-tz}.
  Recent examples of tone-deaf and manipulative behavior
  \cite{Crawford2021-xv,Floridi2004-xk}---and, worse,
  encouragements of self-harm or violence
  \cite{Hill2025-gn,Singleton2023-px}---demonstrate
  significant gaps in AI relative to human emotional intelligence \cite{paech2023eqbench,zhang2025mmeemotion,hu2025emobenchm}.
\end{itemize}

An ideal dataset should comprise each of our three proposed data types:
massively scaled behavioral data that is used to train cognitive models from which the latent variables underlying behavior can be observed;
process-tracing data; and paired neural-behavioral data in the intensive
neuroimaging style. Process supervision and neural data capture
intermediate computations that are invisible in outcome data alone, but
increase the difficulty and cost of data acquisition.

We currently do not have scaling laws needed to predict the scale of
neural data needed for directly training and fine-tuning on neural data.
However, prior experience on large-scale neuroscience datasets, for
example, the Natural Scenes Dataset \cite{Allen2022-bq},
the Allen Brain Observatory Visual Coding dataset
\cite{De_Vries2020-og}, or the International Brain Lab
(IBL) dataset \cite{Findling2025-ix}, indicate that
\textasciitilde500 hours of well-labeled, and accessible data enable
widespread re-use in neuroscience; this is a useful intermediate
milestone before deciding whether to scale up exponentially, as required
by scaling laws.

A potentially useful shortcut could be to leverage games with high
loading on CDM capabilities \cite{Allen2023-ub} that
have been validated by cognitive scientists in massively-scaled rich
behavioral data; for example, chess \cite{Russek2025-yk}, Sea Hero Quest \cite{Spiers2023-hf}, or Diplomacy \cite{paquette2019-diplomacy}. Using these same
environments in process elicitation scenarios and during neural
recordings would allow us to obtain process supervision data and
immediately evaluate the relative value of behavioral versus process data
in training better AI. Indeed, scalably collecting CDM means creating
tasks that a broad range of people can perform over an extended period
of time; games can be designed to heavily load on CDM, while also being
sufficiently intrinsically interesting for people to perform the task.

\section{Discussion}\label{discussion}

The idea that human cognition should inform AI is not new, and it is worth
situating CDM against earlier formulations. Classical approaches, including logic-based
systems and cognitive architectures such as SOAR and ACT-R
\cite{Laird1987-soar,Anderson2004-actr}, sought to replicate cognition by
hand-specifying its processes directly in an architecture. These efforts
illuminated the structure of cognition but did not yield scalable
general-purpose systems. 
Modern efforts to fill in the gaps in the jagged intelligence landscape via novel algorithms and architectures (e.g., JEPA \cite{LeCun2022-jepa}) are being pursued at scale \cite{Heim2026-amilabs}. 
Our proposal differs in kind rather than degree: we do
not advocate encoding metacognition, flexibility, or world models as
architectural primitives. Instead, we treat the cognitive process as a
\emph{training signal} to supervise systems that otherwise
learn and scale by the same mechanisms driving modern AI (i.e., self-supervision and reinforcement learning for transformer-based LLMs). Whether process-level
supervision instills CDM-loaded capabilities is an empirical question, and our
proposed \textasciitilde500-hour milestone is intended to test it
before committing to exponential scaling.

The ``dark matter'' framing also has precedent. Ackerman \cite{ackerman2000domain}
applied it to domain-specific knowledge as an unmeasured contributor to adult
intelligence; Zhu et al. \cite{zhu2020dark} applied it
to the unobserved causal and functional structure, including common sense, intuitive
physics, and the latent ``why'' and ``how'' behind behavior, that purely
data-driven deep learning skips, and argued for a more cognition-oriented AI;
and Bolotta and Dumas \cite{bolotta2022social} applied it
specifically to social interaction, arguing that a solipsistic view of
intelligence leaves this dimension largely unexplored in AI. We share with these
accounts the core diagnosis: that much of what produces intelligent behavior is
not directly visible in the data we typically collect. We
adopt the term deliberately to connect with these precedents while making
explicit the distinct sense in which we use it: not that these capabilities are unknown or unnamed but that they are nearly invisible in the behavioral data that trains and evaluates models. They are dark to the instruments of AI, not to science. Our contribution is not the claim
that easy-to-measure capabilities advance fastest, but the actionable account of
the gap and how to close it through the collection of the right CDM-loaded data.

Measurement has been a crucial ingredient in AI, in two distinct roles: as the source of the training signal, and as the yardstick for evaluation
\cite{Russakovsky2014-rf}. Both are limited in
cognitive dark matter domains, focusing on observable behavior rather
than latent cognitive processes, which has made systematic progress
difficult (for critiques of benchmarks, see \cite{eriksson2025can,zhang2025remembering,raji2021ai}). We propose a research program that leverages large-scale
neural and behavioral data to (i) fine-tune AI models to obtain more
general, less-jagged intelligence and (ii) help us better understand human
cognition. We emphasize the dual value of this research: even if we miss
the mark on building less jagged AI---or if conventional AI research
proceeds sufficiently fast that jaggedness is resolved before
neuroscience can have a major impact---foundational datasets to study
cognitive processes will prove invaluable in better understanding how
human cognition works and building a science of intelligence. The stakes
are high: better agency, calibration, and learning-to-learn are critical
for safer and more reliable AI.

\section{Methods}\label{methods}

\textbf{Chess endgame evaluation.} We prompted frontier models (GPT-5.1-Thinking,
Claude Opus 4.5, Gemini 3.0 Pro) via OpenRouter to create a JavaScript
web application with three mate-in-one chess scenarios, with
drag-and-drop functionality. We ran each query 10 times with default
temperature (1.0) and top-p (1.0) settings. Applications were manually
scored for UI functionality (board rendering, drag-and-drop) and chess
validity (legal positions, valid mate-in-one solutions). Position
validity was verified using the Python chess library. We also ran
interactive sessions with Claude Code where we gave the system iterative
feedback in an attempt to trigger an alternative strategy, but failed to
elicit it.

\textbf{AI survival curve analysis.} $t_{50}$ and $t_{80}$ values for frontier models from Anthropic, OpenAI, and Google were downloaded directly from \href{https://metr.org/}{metr.org}. Each curve is a two-parameter Weibull survival function $S(t) = 0.8^{(t/t_{80})^k}$, normalized so all three cross at $(t_{80}, 0.8)$ and plotted against
$\log_{10}(t/t_{80})$. The shape parameter $k$ was determined as follows: $k = 1$ for the memoryless baseline (constant
hazard), $k \approx 0.68$ for current LLMs from the post-GPT-4 mean $t_{80}/t_{50} = 0.19$ via $k = \ln(0.322)/\ln(0.19)$, and $k \approx 0.36$ for humans from Kwa et al.'s HCAST data ($t_{50} \approx 90$ min, $S(16\text{ h}) \approx 0.20$) \cite{Kwa2025-rf}.

\textbf{Survey of neuroimaging datasets.} We aggregated tasks from 19
intensive neuroimaging datasets identified in Kupers et al.
\cite{Kupers2024-sf} and two additional datasets
\cite{Nakai2020-ol,Jung2025-ij}. Tasks were categorized
according to the three-tier AI capability taxonomy of Liu et al. \cite{Liu2025-es},
cross-referenced against the Cognitive Atlas. Categorization was
performed manually for most datasets; for Individual Brain Charting and
Many-Tasks datasets (\textgreater50 tasks each), we used Claude Opus 4.5
for tagging.

\textbf{Literature analysis.} We analyzed two corpora: the NeuroQuery
dataset \cite{Dockes2020-ci} and all neuroscience
preprints from biorxiv and psyarxiv (Jan 1-Dec 10, 2025;
\textasciitilde12,000 studies). Papers were categorized against the Liu
et al. taxonomy using Claude Sonnet 4.5, extracting at most two primary
cognitive domains per paper; only the top cognitive domain was analyzed.
Papers outside the taxonomy scope (structural/connectomic studies,
reviews, methods papers) were excluded from categorization.

\textbf{AI benchmark analysis.} We referenced the technical reports of
the releases of GPT-5.2, Claude Opus 4.5, and Gemini 3.0 Pro
\cite{OpenAI2025-br,Anthropic2025-ks,GDM2025-nw} to
extract the union of benchmarks used to assess these models'
performance. This resulted in 37 benchmarks in all. We then used GPT-5.2 with high reasoning effort to assign
cognitive functions from Table 1 to each benchmark, determining the
maximum AI tier (L1/L2/L3) probed. We re-ran the analysis 50 times to
assess classification stability.

\bibliography{refs, verified_refs}

\end{document}